\documentclass[a4paper,11pt]{article}
\usepackage{pos}

\def\beq{\begin{equation}}
\def\eeq{\end{equation}}
\def\bea{\begin{eqnarray}}
\def\eea{\end{eqnarray}}

\title{Global analysis of nuclear parton distribution functions}

\author*[a]{Michael Klasen}

\affiliation[a]{Institut f\"ur Theoretische Physik, Universit\"at M\"unster, \\ Wilhelm-Klemm-Stra\ss{}e 9, 48149 M\"unster, Germany}

\emailAdd{michael.klasen@uni-muenster.de}

\abstract{
We review the theoretical foundations, methodological approaches and current status of the determination of nuclear parton distribution functions (PDFs). A large variety of measurements in fixed-target and collider experiments provide increasingly precise constraints on various aspects of nuclear PDFs, including shadowing, antishadowing, the EMC effect, Fermi motion, flavour separation, deuteron binding, target-mass and other higher-twist effects. We give particular emphasis to measurements carried out in proton-lead collisions at the Large Hadron Collider, which have revolutionised the global analysis during the past decade. These measurements include electroweak boson, isolated photon, single inclusive hadron, jet, and heavy-flavour observables.}

\FullConference{31st International Workshop on Deep Inelastic Scattering (DIS2024)\\
 8–12 April 2024\\
Grenoble, France\\}

\begin{document}
\hfill MS-TP-24-15
\maketitle

\section{Introduction}

Nuclear parton distribution functions (PDFs) are an important and dynamic current research topic, as they encode our fundamental understanding of the quark and gluon dynamics in protons (p) and neutrons (n) bound in nuclei (A). They also impact on many other fields in particle and nuclear physics such as global analyses of free proton PDFs, which contain a sizable fraction of nuclear data, the three-dimensional tomography of atomic nuclei, for which they represent important boundary conditions, and the initial-state phase transition to the deconfined quark-gluon plasma, which is believed to be related to gluon saturation and classicalisation. While the evolution of nuclear PDFs with the hard scale $Q$ is calculable in perturbative QCD at next-to-leading order (NLO) and beyond, their dependence on the longitudinal parton momentum fraction $x$ must be fitted to experimental data. For the universality of PDFs in different collision processes, one relies on QCD factorisation theorems. Nuclear modifications such as shadowing at $x\lesssim0.05$, antishadowing at $0.05\lesssim x\lesssim0.3$, the famous EMC effect at $0.3\lesssim x\lesssim0.7$, and Fermi motion at $0.7\lesssim x$ have been known for many years and can be parameterised and fitted, but they still remain to be fundamentally understood.

The key process in any PDF analysis is deep-inelastic scattering (DIS) of charged leptons, which for nuclei has so far only been performed in fixed-target (FT) experiments at relatively small energy and which is at LO only sensitive to quarks. Neutrino scattering provides additional constraints on their flavour separation, in particular on the strange quark and in processes with an identified final-state charm meson. The gluon density is directly accessible in hadronic collisions not only in FT experiments, but also at colliders such as BNL's RHIC and CERN's LHC, and the LHC experiments have collected an enormous amount of data in pPb collisions during the last decade. Since higher-twist (HT) effects, e.g.\ from final-state rescattering, are enhanced in nuclear collisions, a sufficiently large scale is required to ensure a leading-twist (LT) interpretation of the data. The Fermi motion of individual nucleons seems to be physically reasonably clear, but for shadowing both LT and HT explanations have been proposed, for the EMC effect both partonic and hadronic mechanisms, and antishadowing is mainly justified with the momentum sum rule. More details can be found in our recent review of nuclear PDFs and the references cited therein \cite{Klasen:2023uqj}. Here we present a concise summary of this review, but provide also some complementary information.

\section{Methodology}

In deep-inelastic scattering, the nuclear structure function
\bea
 F_2^A(x,Q^2)&=&\sum_if_i^{(A,Z)}(x,Q^2)\otimes C_{2,i}(x,Q^2)
\eea
factorises into perturbatively calculable Wilson coefficients $C_{2,i}$, that are currently known at next-to-next-to-leading order (NNLO) and partially beyond \cite{Falcioni:2024xyt}, and nuclear PDFs
\bea
 f_i^{(A,Z)}(x,Q^2)&=&{Z\over A}f_i^{p/A}(x,Q^2)+{A-Z\over A}f_i^{n/A}(x,Q^2),
\eea
which can be technically split into (unphysical) bound nucleon PDFs. Here, $A$ and the $Z$ denote the nuclear mass and charge. Evolution of the PDFs with $Q^2$ is governed by the DGLAP equations
\bea
 {\partial f_i(x,Q^2)\over\partial\log Q^2}&=&\int_x^1{dz\over z}P_{ij}\left({x\over z},\alpha_s(Q^2)\right)f_j(z,Q^2)
\eea
with perturbative splitting functions $P_{ij}$. In addition, the PDFs have to satisfy number and momentum rules. Furthermore, for nuclear PDFs isospin symmetry
\bea
 f^{n/A}_{d,u}(x,Q^2)&=&f^{p/A}_{u,d}(x,Q^2)
\eea
is usually assumed. Prior to BNL's RHIC and CERN's LHC, global nuclear PDF analyses relied on neutral-current (NC) and charged-current (CC) DIS and Drell-Yan (DY) FT data and still do so for nuclei other than Au (RHIC) and Pb (LHC). This is, however, about to change with Xe and O in the LHC and in particular with BNL's EIC \cite{AbdulKhalek:2022hcn}.

Theoretically, the global nuclear PDF analyses nCTEQ15HQ \cite{Duwentaster:2022kpv}, EPPS21 \cite{Eskola:2021nhw} and nNNPDF3.0 \cite{AbdulKhalek:2022fyi} have so far been performed only at NLO due to the unavailability of fast NNLO codes for processes other than DIS, DY and W- and Z-boson production, to which the KSASG20 \cite{Khanpour:2020zyu} and TUJU21 \cite{Helenius:2021tof} NNLO analyses were therefore restricted. In DGLAP evolution codes, heavy-quark (HQ) masses are usually taken into account with general-mass variable flavour number schemes such as FONLL, but extensions to the SACOT-$\chi$ scheme at NNLO including charged-current DIS are currently in progress \cite{Risse:2023rxd}. Fitting of the experimental data is performed by optimising the $\chi^2$ figure-of-merit function and propagation of the experimental uncertainties into the PDFs with the Hessian or Monte Carlo method. The validity of the Hessian approximation and the interpretation of the Monte Carlo uncertainty receive currently a lot of interest (cf.\ the contributions by M.\ Costantini, N.\ Derakhshanian, T.\ Giani and P.\ Risse at this workshop) \cite{Derakhshanian:2023bed}.

Nuclear data taken with deuterons (D) in FT experiments and at RHIC must be corrected for Fermi motion, weak binding and off-shell effects, in particular at intermediate and large values of $x$ (cf.\ the contributions by W.\ Henry and M.\ Cerutti) \cite{Accardi:2016qay}. The dominant target mass corrections (TMCs) can be taken into account with the Nachtmann scaling variable $\xi=2x/(1+r)$, where $r=\sqrt{1+4x^2M^2/Q^2}$, and they depend only weakly on $A$. Subleading TMCs contribute at HT (cf.\ the contribution by R.\ Ruiz) \cite{Ruiz:2023ozv}. At high $x$ and low $Q$ values, HT corrections can generally become important and can be parameterised and fitted additively or multiplicatively (cf.\ the contributions by M.\ Cerutti and R.\ Petti). Since many neutrino DIS data points lie in this kinematic region, all of these corrections have to be taken into account there \cite{Segarra:2020gtj}. Unfortunately, they cannot fully resolve the long-standing tension of the CC with the NC DIS data, so that in practice only a subset of neutrino data is used for the important separation of quark flavours (cf.\ the contributions by J.\ Rojo and S.\ Yrjänheikki) \cite{Muzakka:2022wey}. Much hope lies therefore on future neutrino DIS data from the FASER$\nu$ and SND@LHC experiments at the LHC Forward Physics Facility (cf.\ the contributions by J.\ Atkinson and O.\ Durhan). With electroweak boson, isolated photon, hadron and jet as well as heavy-quark and quarkonium data, the four main LHC experiments have already extended the kinematic plane by orders of magnitude and down to values of $x<10^{-5}$ and up to $Q^2>10^5$ GeV$^2$.

\section{Electroweak bosons}

\begin{figure}
 \centering
 \includegraphics[width=\textwidth]{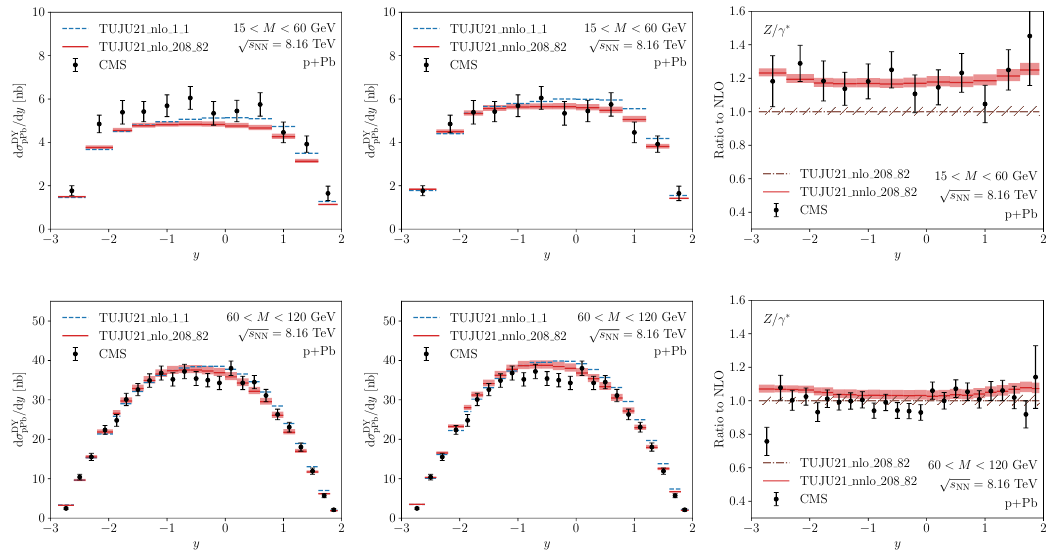}
 \caption{Comparison of DY production in pPb collisions at $\sqrt{s} = 8.16$ TeV at NLO (left) and NNLO (center) results with (solid with uncertainty band) and without (dashed) nuclear PDF modifications in two invariant mass bins, $15 < M < 60$ GeV (upper panels) and $60 < M < 120$ GeV (lower panels) to CMS data \cite{CMS:2021ynu}. In the right part we plot the ratios of the NNLO (red with uncertainty) and NLO (dot-dashed brown with hatched uncertainty) together with the data. Taken from Ref.\ \cite{Helenius:2021tof}.}
 \label{fig:1}
\end{figure}

The analysis of $W$- and $Z$-boson production in pPb collisions at LHC Runs I and II has been led by CMS and ALICE and is now seconded by LHCb and ATLAS. The importance of these data lies in their potential sensitivity to flavour separation and the strange quark, which in practice is, however, smaller than naively expected due to the radiative generation of strange quarks from gluon splittings \cite{Kusina:2020lyz}. DY and W-/Z-boson production thus also impose constraints on the gluon density \cite{Berger:1998ev,Klasen:2013ulb,Brandt:2014vva,Andronic:2024rfn}. Particularly precise are the Run-II $Z$-boson data from CMS shown in Fig.\ \ref{fig:1} for two invariant-mass windows and as a function of the rapidity $y$ of the dilepton pair \cite{CMS:2021ynu}.
A comparison of these CMS data with NLO and NNLO calculations using TUJU21 PDFs shows that the NNLO corrections are rather mild (around 5\%) for the high-mass bin, but become significant for the low-mass bin, reaching 20\% at the largest (absolute) rapidities \cite{Helenius:2021tof}. Unfortunately, large fluctuations at mid-rapidity make it difficult to have acceptable $\chi^2$-values even at NNLO. In contrast, the CMS Run-II $W$-boson data \cite{CMS:2019leu} have been used in TUJU21 and all three global NLO analyses, with nCTEQ15HQ and nNNPDF3.0 fitting absolute cross sections and EPPS21 ratios of cross sections.

\section{Photons, hadrons and jets}

\begin{figure}
 \includegraphics[width=\textwidth]{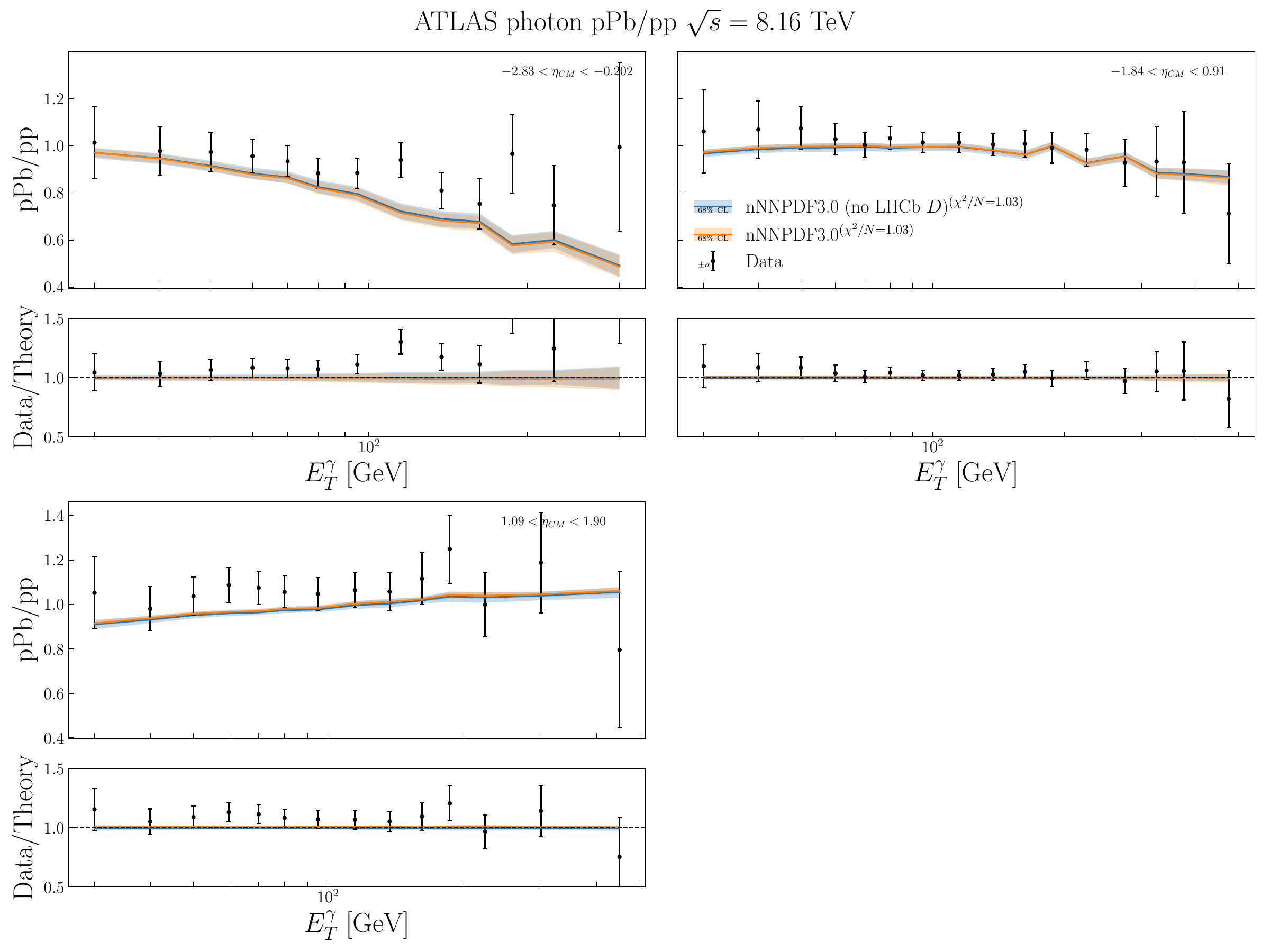}
 \caption{Comparison between ATLAS pPb/pp prompt photon cross section ratios \cite{ATLAS:2019ery} and theoretical predictions based on nNNPDF3.0 PDFs fitted with and without LHCb D-meson data, differential in the photon transverse energy ($E_T^\gamma$) and for three photon pseudo-rapidity ($\eta_{\rm CM}$) bins. The bands in the theory predictions indicate the PDF uncertainty. The bottom panels display the ratio to the central theory. The values of $\chi^2$ to this dataset for both nNNPDF3.0 fits are indicated in the legend. Taken from Ref.\ \cite{AbdulKhalek:2022fyi}.}
 \label{fig:2}
\end{figure}

Prompt photon production is well-known to be sensitive to the gluon density through the QCD Compton process \cite{Klasen:2017dsy}. Prior to the LHC, prompt photon data have been taken by FNAL's FT experiment E706 in pBe and by the PHENIX and STAR experiments at RHIC in DAu collisions. The only published LHC pPb cross sections, taken differentially in the photon transverse energy ($E_T^\gamma$) and for three photon pseudo-rapidity ($\eta_{\rm CM}$) bins at $\sqrt{s}=8.16$ TeV, come from ATLAS \cite{ATLAS:2019ery}. Their ratios to ATLAS pp cross sections, taken with the same isolation criterion and extrapolated from $\sqrt{s}=8$ TeV \cite{ATLAS:2016fta}, are shown in Fig.\ \ref{fig:2}.
The data are compared with NLO predictions based on nNNPDF3.0 PDFs fitted with and without LHCb D-meson data. The bands in the theory predictions indicate the PDF uncertainty. For pPb/pp ratios, a good description of the data can be achieved at NLO, and they have therefore been included in the nNNPDF3.0 fit \cite{AbdulKhalek:2022fyi}. However, the absolute cross sections are underestimated at NLO, which may indicate a need for NNLO precision. Preliminary ALICE data \cite{jonas} are consistent with the ATLAS measurements, but extend them to lower $E_T^\gamma$ and thus lower $x$ values in the nuclear PDFs, so that their publication is eagerly awaited.


Single inclusive hadron (SIH, specifically $\pi^0$) production at RHIC was one of the first processes used to impose constraints on the gluon density. In the meantime, many more data sets, also on charged pions, kaons, and $\eta$ mesons, have become available from PHENIX and STAR at RHIC and in particular ALICE at the LHC and have been used to extend the original nCTEQ15 analysis \cite{Kovarik:2015cma} to nCTEQ15SIH \cite{Duwentaster:2021ioo}. As the SIH calculation depends on the fragmentation of the final parton into the observed hadron, a variety of fragmentation functions available in the literature was used to properly estimate this source of uncertainty. This was not possible for $\eta$ mesons, which, however, also had only a minor impact on the nCTEQ15SIH analysis.

\begin{figure}
 \includegraphics[width=0.49\linewidth]{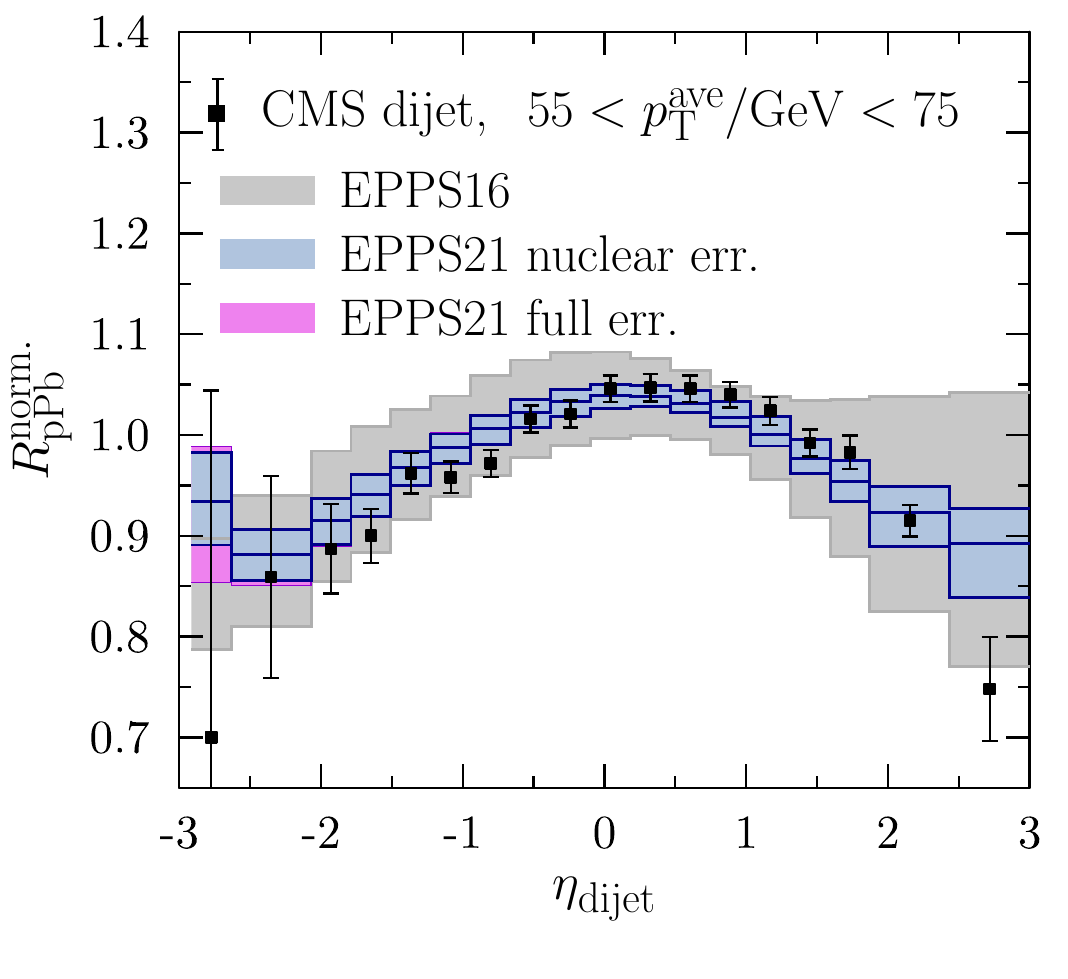}
 \includegraphics[width=0.49\linewidth]{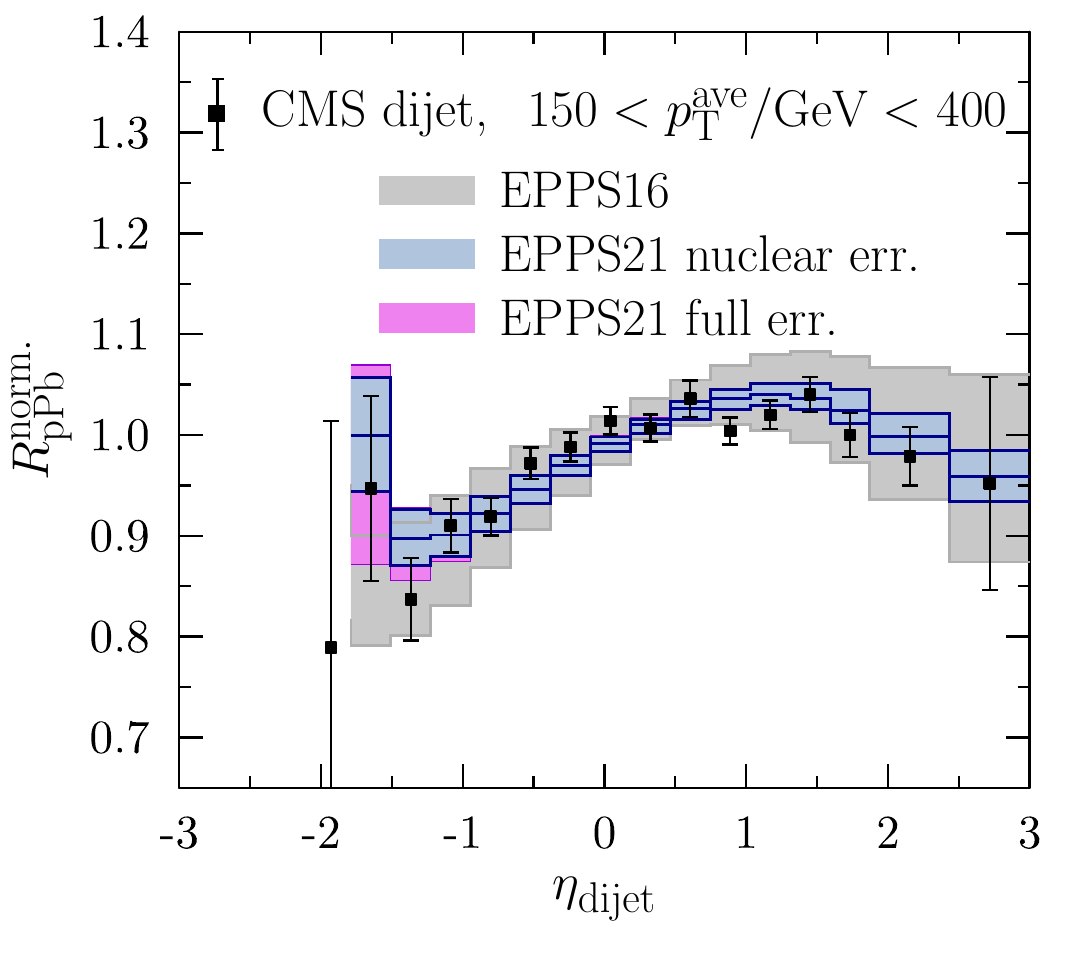}
 \caption{The CMS dijet data \cite{CMS:2018jpl} compared with the EPPS21 analysis. The solid blue lines show the central results, inner blue bands the nuclear uncertainties, and the purple bands the total uncertainty. The grey bands correspond to the EPPS16 results \cite{Eskola:2016oht}. Taken from Ref.\ \cite{Eskola:2021nhw}.}
 \label{fig:3}
\end{figure}

Hadronic jets are less sensitive to details of the fragmentation process than single inclusive hadrons. In nuclear collisions one must, however, pay particular attention to the background from the so-called underlying event, which is much larger than in pp collisions and has been estimated to involve on average $7\pm5$ pN interactions \cite{Loizides:2017ack}. Multiparton scatterings are therefore usually modelled by Monte Carlo simulations and subtracted from the experimental data. For further reduction of the associated uncertainty, one resorts to sufficiently large transverse momenta and small jet cones. Currently, the most constraining data have been taken in LHC Run I by CMS on dijets differential in the average $p_T$ ($p^{\rm ave}_{\rm T}$) and rapidity ($\eta_{\rm dijet}$) of the two jets \cite{CMS:2018jpl}. They supersede earlier dijet data, which were included in the EPPS16 analysis \cite{Eskola:2016oht}. The cross sections are normalised to the rapidity-integrated cross section, so that most of the systematic uncertainties cancel. The resulting spectra in pp are then so precise that they challenge the theoretical description, the NLO perturbative QCD calculations being in tension with the data \cite{Eskola:2019dui}. The impact of the NNLO corrections \cite{Currie:2017eqf} is still unclear. Despite this tension, ratios of normalised cross sections between pPb and pp collisions, $R_{\rm pPb}^{\rm norm.}$, are broadly consistent with nuclear PDFs. They have therefore been included in the EPPS21 \cite{Eskola:2021nhw} and nNNPDF3.0 \cite{AbdulKhalek:2022fyi} global fits, where they have a large impact on the gluon. Fig.\ \ref{fig:3} shows the large reduction of the nuclear uncertainty in the forward region from EPPS16 (grey) to EPPS21 (blue), while in the backward region the gluon uncertainty in the underlying CT18A proton PDFs \cite{Hou:2019efy} adds to the full error (purple). However, the most forward data points at the edge of the detector acceptance indicate a suppression, which cannot be fitted.

\section{Heavy quarks and quarkonia}

\begin{figure}
 \includegraphics[width=0.7\textwidth]{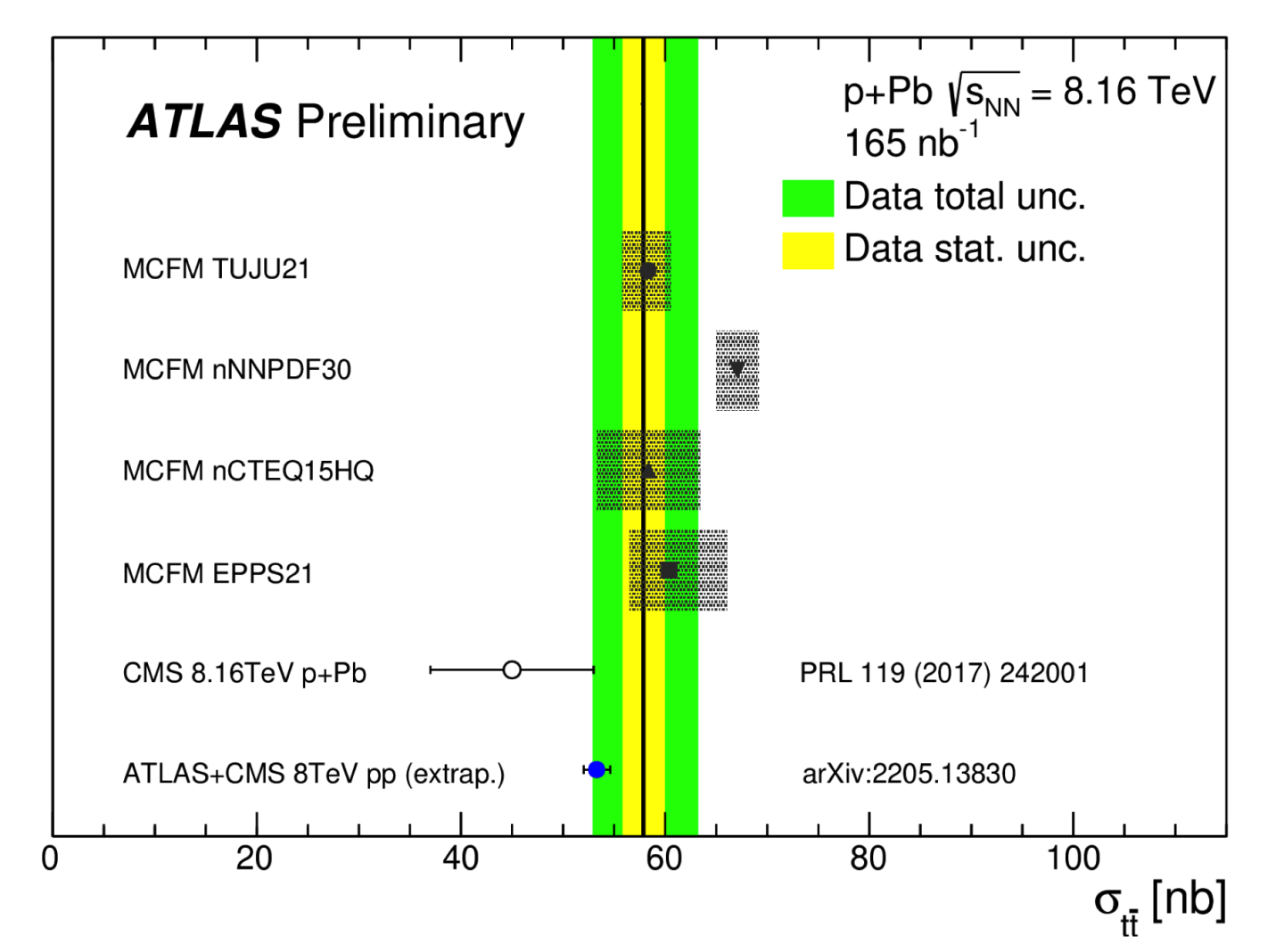}
 \centering
 \caption{Comparison of the observed top quark cross section in ATLAS \cite{ATLAS:2023oce} with the corresponding CMS measurement \cite{CMS:2017hnw} at $\sqrt{s}= 8.16$ TeV. Also shown is the combined ATLAS and CMS measurement in pp collisions at $\sqrt{s}=8$ TeV, extrapolated to 8.16 TeV. Theoretical predictions are calculated at NNLO for different nuclear PDF sets with associated PDF uncertainties (grey). The central ATLAS measurement is indicated by the black line, the statistical uncertainty and its combination with the systematic uncertainty by yellow and green shaded areas. Taken from Ref.\ \cite{ATLAS:2023oce}.}
 \label{fig:4}
\end{figure}

During the last decade, the four LHC experiments and in particular ALICE and LHCb have taken vast data sets on open heavy quark and quarkonium production. They cover a wide kinematic range and put strong constraints on the nuclear gluon PDF down to $x\leq 10^{-5}$. Theoretical predictions for these data sets can be obtained from a data-driven approach, where pp data are used to determine effective scattering matrix elements. Including an explicit rapidity dependence in the ansatz allows to describe also the very forward LHCb data. This approach has been validated with detailed comparisons to existing NLO calculations in non-relativistic QCD (NRQCD) for quarkonia and in the general-mass variable-flavor-number scheme (GM VFNS) for the open heavy-flavoured mesons (cf.\ the contribution by J.\ Wissmann). Inclusion of these data has led to an enormous reduction in the gluon uncertainty in the nCTEQ15HQ analysis \cite{Duwentaster:2022kpv}. In addition, the uncertainties from the data-driven approach have been determined using the Hessian method and accounted for in the PDF fits (cf.\ the contribution by T.\ Jezo). One of the most recent LHC measurements in pPb collisions is the observation of top quark production by ATLAS \cite{ATLAS:2023oce} and CMS \cite{CMS:2017hnw}. The measured cross sections are compared to the PDFs discussed in this contribution in Fig.\ \ref{fig:4} and show for most of them an impressive agreement, even though these data were not included in the fits.

\section{Conclusion}

\begin{figure}
 \includegraphics[width=\textwidth]{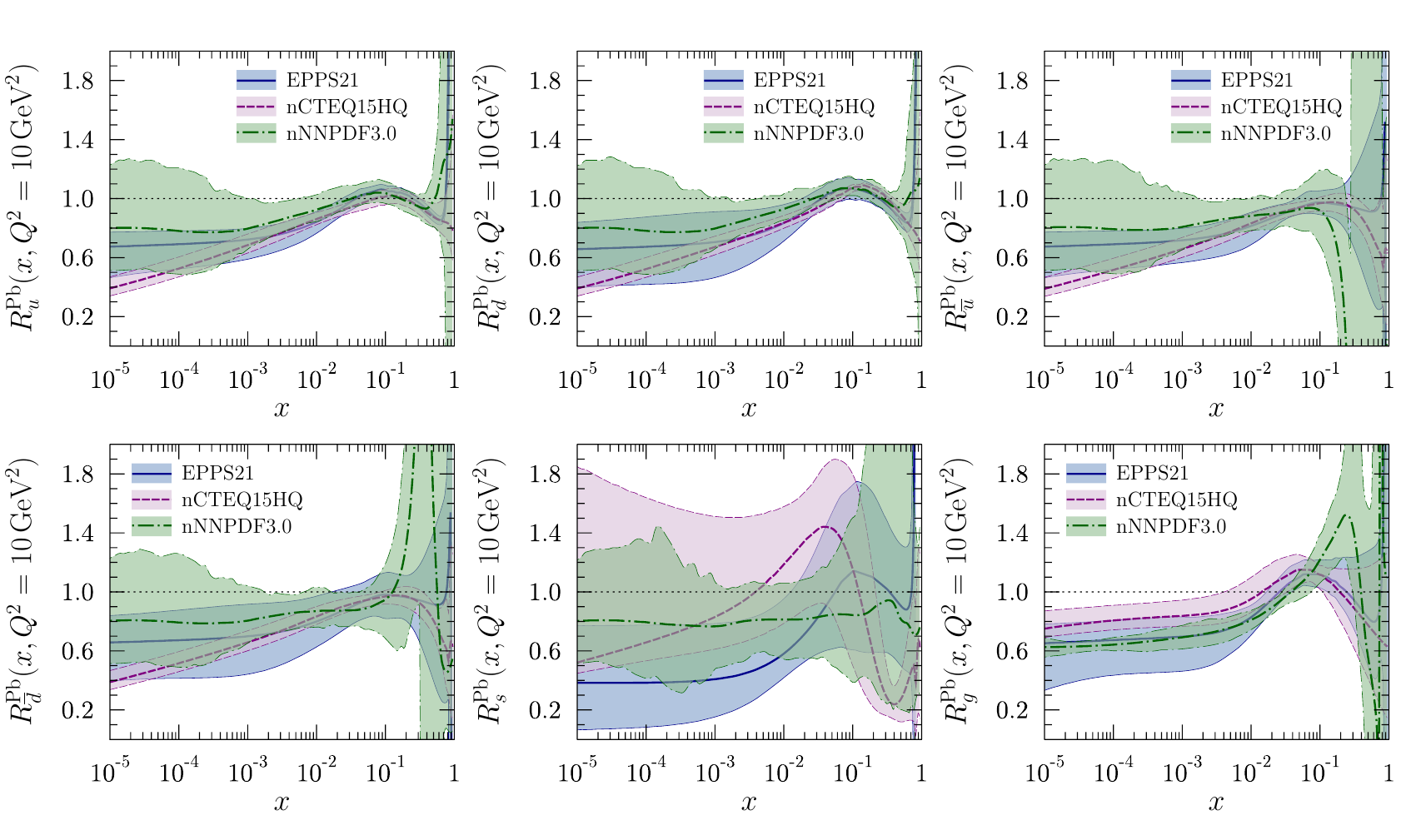}
 \caption{Comparison of the $^{208}$Pb nuclear modifications resulting from the EPPS21 (full, blue) \cite{Eskola:2021nhw}, nCTEQ15HQ (dashed, red) \cite{Duwentaster:2022kpv} and nNNPDF3.0 (dot-dashed, green) \cite{AbdulKhalek:2022fyi} global analyses of nuclear PDFs, i.e.\ the PDFs of lead divided by the summed PDFs of 82 free protons and 126 free neutrons. Uncertainty bands correspond to 90\% CL. Taken from Ref.\ \cite{Klasen:2023uqj}.}
 \label{fig:5}
\end{figure}

To summarise, Fig.\ \ref{fig:5} compares the nuclear modifications of the lead nucleus PDFs at $Q^2 = 10$ GeV$^2$ from EPPS21 (full, blue) \cite{Eskola:2021nhw}, nCTEQ15HQ (dashed, red) \cite{Duwentaster:2022kpv} and nNNPDF3.0 (dot-dashed, green) \cite{AbdulKhalek:2022fyi}. Qualitatively, there is good overall agreement between all three within the 90\% CL uncertainty bands (shaded areas). Closer inspection reveals nevertheless still significant differences both among the central values and the widths of the uncertainty bands in several distributions and $x$ regions. The vastly expanding measurements at the LHC and steady theoretical progress, including recent advances in lattice QCD calculations (cf.\ the contribution by H.W.\ Lin), will ensure continuous progress in the important dynamic field of nuclear PDFs over the next years, as we eagerly await the commissioning of BNL's EIC.

\section*{Acknowledgment}

The author thanks the organisers of this workshop for the kind invitation and his nCTEQ colleagues as well as H.\ Paukkunen for their collaboration. This work has been supported by the BMBF under contract 05P21PMCAA and by the DFG through CRC ``Isoquant'', project-id 273811115, as well as RTG 2149 ``Strong and Weak Interactions - from Hadrons to Dark Matter''.

\end{document}